\documentclass[11pt,twoside]{article}
\usepackage{asp2006}
\usepackage{psfig}
\usepackage{epsf}
\usepackage{graphicx}
\usepackage{lscape}
\def\edcomment#1{\iffalse\marginpar{\raggedright\sl#1\/}\else\relax\fi}
\reversemarginpar

\begin{document}
\title{Observations of Supersonic Downflows in a Sunspot Light Bridge as Revealed by {\em Hinode}}
\author{Rohan E. Louis}
\affil{Udaipur Solar Observatory, Physical Research Laboratory
                     Dewali, Badi Road, Udaipur,
	       	     Rajasthan - 313004, India}
\author{Luis R. Bellot Rubio}
\affil{Instituto de Astrof\'{\i}sica de Andaluc\'{\i}a (CSIC),
                     Apartado de Correos 3004,
                     18080 Granada, Spain}
\author{Shibu K. Mathew, P. Venkatakrishnan}
\affil{Udaipur Solar Observatory, Physical Research Laboratory
                     Dewali, Badi Road, Udaipur,
	       	     Rajasthan - 313004, India}

\begin{abstract}
Recent high resolution spectropolarimetric observations from Hinode 
detected the presence of supersonic downflows in a sunspot light bridge \citep{Rohan2009}.
These downflows occurred in localized patches, close to regions where 
the field azimuth changed by a large value. This apparent discontinuity 
in the field azimuth was seen along a thin ridge running along the western 
edge of the light bridge. Some, but not all, of these downflowing patches were co-spatial 
with chromospheric brightness enhancements seen in Ca {\sc ii} H filtergrams. The 
presence of magnetic inhomogeneities at scales of 0.$''$3 could 
facilitate the reconnection of field lines in the lower chromosphere 
whose signatures might be the supersonic downflows and the 
brightness enhancements that have been observed.

\end{abstract}

\vspace{-0.5cm}
\section{Introduction}
\label{intro}
Sunspot Light Bridges (LBs) are conspicuous bright intrusions in the 
otherwise cool, dark umbra and are often seen to exhibit a granular 
like morphology, bearing a ubiquitous central dark lane 
\citep{Hirzberger2002}. LBs are known 
to harbour weak inclined magnetic fields 
\citep{Ruedi1995,Leka1997,Jurcak2006,Katsukawa2007} but the 
nature of sub-photospheric convection that powers them is still 
a matter of debate \citep{Parker1979,Spruit2006,Rimmele2004}. 
The presence of supersonic downflows in a sunspot LB in NOAA 10953,
using high resolution spectropolarimetric observations from 
{\em Hinode} \citep{Kosugi2007} on May 1, 2007, also indicate chromospheric enhancements 
that are co-spatial with some of the strong downflowing patches \citep{Rohan2009}. 
In this paper, we show that reconnection in the lower chromosphere 
could be a possible mechanism that could explain the downflows and the 
chromospheric activity.

\section{Results}
\label{res}

\subsection{Supersonic Downflows}
\label{downflows}
\begin{figure}[!h]
\centerline{
\includegraphics[width=0.4\textwidth,angle=90]{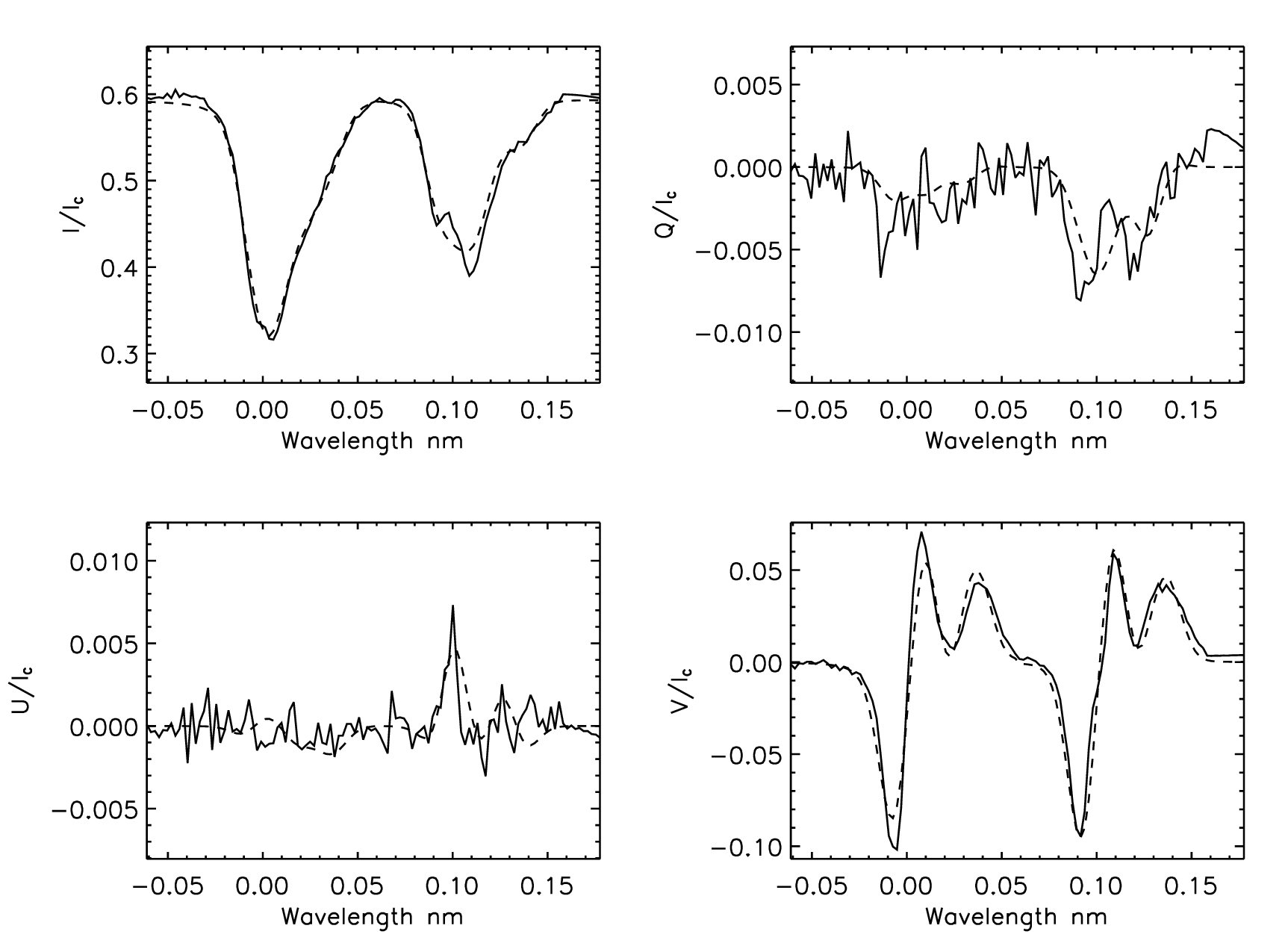}
\includegraphics[width=0.4\textwidth,angle=90]{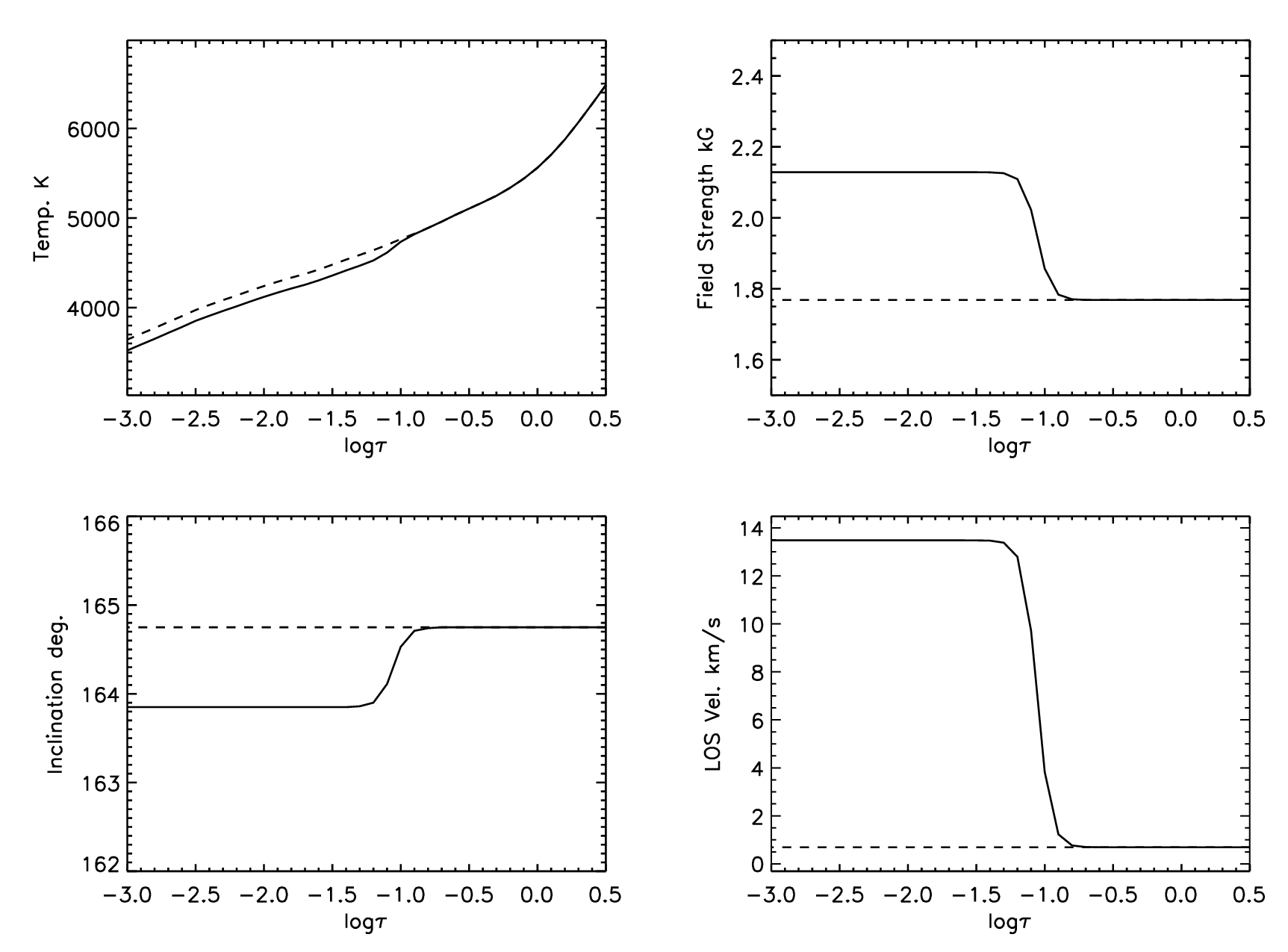}
}
\caption{{\bf{First and second columns: }}Observed ({\em solid}) and best-fit ({\em dashed}) Stokes profiles using a two-component model in which one of the components has a discontinuous stratification. {\bf{Third and fourth columns: }}The constant background atmosphere is represented by the dashed lines, whereas the solid lines correspond to the discontinuous atmosphere.  The supersonic downflows are seen in the upper half of the photosphere.}
\label{inv_results3}
\end{figure}

The strong downflows were observed as localized patches with velocities of $\approx$
4 km.s$^{-1}$ (See Figure 1 of \citet{Rohan2009}), that were retrieved from the 
SIR inversion code \citep{Ruiz1992}. The Stokes profiles associated with the strong 
downflows comprise of two peaks in the red lobe which cannot be reproduced by a simple
model atmosphere where all parameters are constant with height. In order to synthesize such
anomalous profiles, a two-component model atmosphere is used as shown in Figure~\ref{inv_results3}, 
where the components are stacked one on top of another. The amplitude and location of the
discontinuity in the stratification are free parameters in the inversion code (SIRJUMP; Bellot Rubio in preparation).
The anomalous profiles are well reproduced by this model atmosphere and as seen from the 
Figure, there exist supersonic downflows in the upper half of the photosphere. \citet{Rohan2009} had
shown that supersonic velocities are also retrieved by other model atmospheres with two-components.

\begin{figure}[!h]
\centerline{
\includegraphics[width=0.6\textwidth,angle=90]{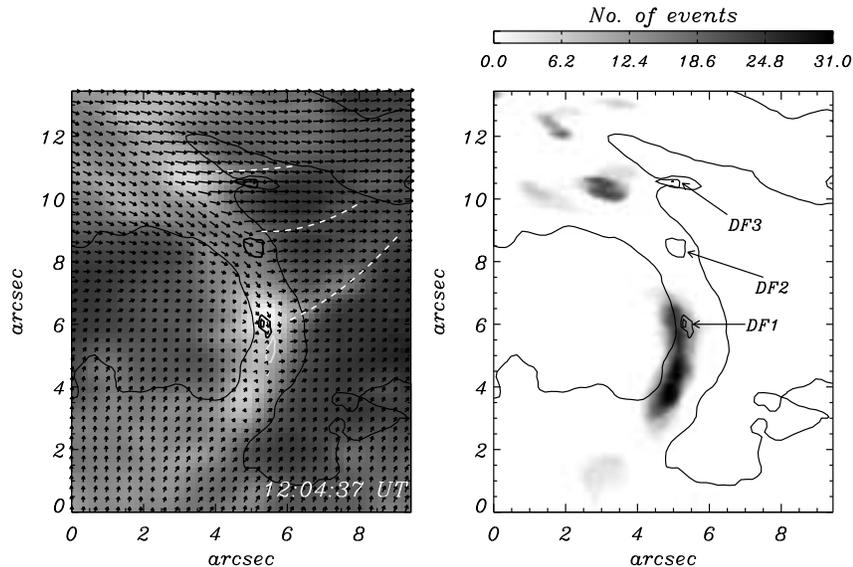}
}
\vspace{-10pt}
\caption{{\bf{Left: }}Ca {\sc ii} H filtergram of the LB taken at 12:04:37 UT when the spectrograph slit was above the LB. The arrows indicate the transverse component of the field in the local reference frame for every alternate pixel. White contours represent fields weaker than 750~G and black contours outline bright continuum structures. The white dashed lines mark thin chromospheric threads whose ends are located in or near the downflowing patches. {\bf{Right: }}Chromospheric event map constructed during the polarimetric scan of the LB and its neighbourhood. DF1, DF2 and DF3 indicate the downflowing regions. Black Contours have been drawn for 2.5 and 4 km.s$^{-1}$ respectively in both the panels.}
\label{calcium}
\end{figure}

\subsection{Chromospheric Activity}
\label{chromosphere}
Figure~\ref{calcium} shows that some, but not all, of the strong downflowing
patches coincide with chromospheric Ca {\sc ii} H brightness enhancements.
During the early part of May 1, \citet{Rohan2008} had observed that the 
chromosphere above the LB exhibited transient flashes across 
it, as well a strong brightening along a loop-like structure that 
was seen in the 3 hr mean image. It was also shown that the transient 
events were extremely fast and had lifetimes of 1-2 min. These enhancements
were also observed on April 29 and 30 \citep{Shimizu2009}. 
In order to study the association of the downflows to the chromospheric 
enhancements, events maps were constructed in the following manner.
All Ca {\sc ii} H filtergrams in the 44 min. sequence were normalized by the Quiet 
Sun intensity. Using an intensity threshold of 0.9, a binary image was created for each 
individual image in the time sequence, setting all pixels greater than the 
threshold to one and the rest to zero. These binary maps were then added 
in time, yielding at the end, a map with pixels having values indicating the number 
of chromospheric events. The right panel of Figure~\ref{calcium}
illustrates the event map constructed from the manner described 
above. The strongest downflowing patch DF1 is seen, lying within the sausage shaped 
region where a large number of events occurred. The two patches DF2 and DF3 lie
$\approx$ 2$''$ and 3$''$ away from the nearest chromospheric brightening.

\section{Discussion}
\label{discuss}
The supersonic downflows in the LB are sometimes associated with 
transient chromospheric brightenings. This is illustrated from the chromospheric
event map. The question that naturally arises from the above observations is 
the source/mechanism of the supersonic downflows and the brightenings? The strong 
downflows are similar to the Evershed Flow which can be supersonic in the mid 
and outer penumbra as well as beyond the sunspot boundary \citep{Josecarlos2001,Luis2004}. Although, the flow 
mechanisms responsible for Evershed Flow \citep{Montesinos1997,Schlichenmaier1998} cannot be ruled out as one 
possibility for the supersonic downflows observed in the LB, it seems unlikely 
that it would account/explain the associated chromospheric phenomena as well. 
In the LB, of the three downflowing patches, two of them on the upper 
half of the LB are associated with azimuth discontinuities \citep{Rohan2009}, which leads us 
to believe that a slingshot reconnection mechanism, as formulated by \citet{Ryutova2008}, could 
produce the downflows as well as explain the chromospheric brightenings. The 
question is where does the reconnection occur? One useful hint lies in 
the inclination of the magnetic field. The top most 
downflowing patch (DF3) on the LB is $\approx$ 2$''$ or 1450 km, from 
the nearest chromospheric brightening. The inclination of the magnetic 
field is $\approx$ 120$^{\circ}$ to the vertical. If a reconnection geometry 
as suggested by \citet{Isobe2007} is assumed, then the 
height at which the reconnection occurs would be 1450$\times\tan{30}^{\circ}$ km
or $\approx$ 840 km which is well within the formation height of the Ca {\sc ii} H
line. By the same argument, it can be shown that if reconnection occurs at this 
height, then for fields which have an inclination of 165$^{\circ}$, the distance 
between the chromospheric brightening and the downflows would be $\approx$ 225 km 
or 0.3$''$, which is 2 SP pixels. This is precisely the case for the 
strongest downflowing patch (DF1) that is co-spatial with the chromospheric 
brightenings. The chromosphere can facilitate reconnection as plasma $\beta$ is 
less than unity at these heights. Thus the photospheric downflows could be the result of
downward propagating shocks resulting from the reconnection. 

\acknowledgements We sincerely thank the Hinode team for providing the 
high resolution data. Hinode is a Japanese mission developed and launched by
ISAS/JAXA, with NAOJ as domestic partner and NASA and
STFC (UK) as international partners. It is operated by these
agencies in co-operation with ESA and NSC (Norway).

\end{document}